\documentclass[letter]{jpsj2} %% for letters
\title{What is Minimal Model of $^3$He Adsorbed on Graphite? \\ --Importance of Density Fluctuations in 4/7 Registered Solid--}

\author{Shinji \textsc{Watanabe} and Masatoshi \textsc{Imada}}

\inst{Department of Applied Physics, University of Tokyo, Hongo 7-3-1, Bunkyo-ku, Tokyo, 113-8656, Japan}

\abst{
We show theoretically that the second layer of $^3$He adsorbed on graphite and solidified at 4/7 of the first-layer density is close to the fluid-solid boundary with substantial density fluctuations on the third layer. The solid shows a translational symmetry breaking as in charge-ordered insulators of electronic systems.
We construct a minimal model beyond the multiple-exchange Heisenberg model.  
An unexpectedly large magnetic field required for the measured saturation of magnetization is well explained by the density fluctuations.  The emergence of quantum spin liquid is understood from the same mechanism as in the Hubbard model and in $\kappa$-$\rm (ET)_2Cu_2(CN)_3$ near the Mott transitions. 
}

\kword{quantum spin liquid, $^3$He, 4/7 phase, saturation field, Mott insulator, charge order}

\begin{document}
\maketitle

%\section{Introduction}
$^3$He layers adsorbed on graphite substrate is a unique two-dimensional correlated Fermion
system and have continuously offered fundamental issues in condensed matter.  In particular, adsorption of $^3$He to the 2nd layer under the corrugation potential of the 1st-layer solid shows a variety of phenomena ranging from a correlated Fermi liquid to a solidification at the commensurate density of 4/7 relative to the 1st layer~\cite{Elser}.  The solid phase bahaves as a quantum spin liquid (QSL)~\cite{masutomi,ishida}, the nature of which is a long-standing theoretical challenge~\cite{Anderson}.  

This solidified $^3$He monolayer has widely been studied by the Heisenberg model with multiple spin exchange (MSE)~\cite{Roger,ikegami}.
However, exact diagonalization studies on realistic MSE models suggest an opening of spin excitation gap~\cite{misguich} in contrast to the gapless nature of QSL revealed by specific heat~\cite{ishida,tuji} and magnetic susceptibility measured down to 
10 $\mu$K~\cite{ikegami,masutomi}.  Furthermore, the MSE model predicts that the magnetization $m$ saturates above the field $h_{\rm sat}\sim 7$ Tesla~\cite{misguich,momoi}, 
whereas a recent experiment~\cite{ishimoto} 
up to 10 Tesla indicates the saturation at much higher $h_{\rm sat}$.

A gapless QSL was reported in numerical studies as the ground state of the two-dimensional Hubbard model with geometrical frustration effect near the Mott transition~\cite{KI,MWI,Mizusaki} supplemented by a report showing numerically the absence of various symmetry breakings~\cite{Watanabe}. It supports the realization of a genuine Mott insulating state without any translational symmetry breaking as initially proposed by Anderson~\cite{Anderson}.   This series of studies is indeed relevant and provides a realistic model for a subsequently discovered gapless spin liquid in $\kappa$-(ET)$_2$Cu$_2$(CN)$_3$.  Although a charge gap exists in Mott insulating states, density fluctuations allowing doubly occupied sites in the Hubbard model near the Mott transition is crucial for the stabilization of the QSL. 

However, when we consider the hard core of the interatomic interaction between $^3$He atoms, the Hubbard model with a moderate onsite interaction $U$ near the Mott transition with a crucial role of density fluctuations looks unrealistic as a model of $^3$He monolayer.

In this letter, we show that the 4/7-density solid is actually located in the vicinity of the fluid-solid boundary implying essentially the same character as the QSL found in the Hubbard model with substantial density fluctuations, contrary to the conventional picture.  More precisely, the density fluctuation in the solid between the 2nd and 3rd layers accompanied by a translational symmetry breaking on the 2nd layer solves the puzzles: It causes 
enhancement of the ratio of $h_{\rm sat}$ to the exchange interaction as is revealed in the recent experiment~\cite{ishimoto}. Furthermore, it naturally explains why the MSE model is insufficient to describe the 4/7 phase of $^3$He. 

\begin{figure}[t]
\begin{center}
\includegraphics[width=7cm]{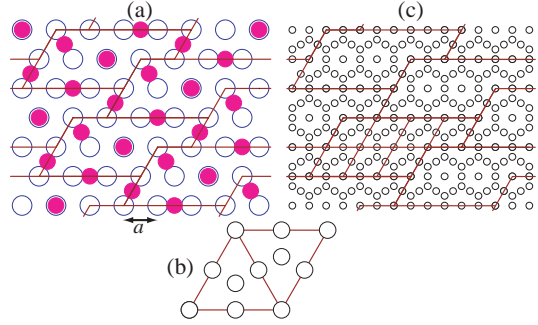}
\end{center}
\caption{
(Color online) 
(a) Lattice structure of the 4/7 phase of $^3$He. 
Both 1st-layer atoms (open circles) and 2nd-layer atoms (shaded circles) 
form triangular lattices in the solid phase. 
The area enclosed by the solid line represents the unit cell for the solid 
of the 2nd-layer atoms (see text). 
The lattice constant of the 1st layer is $a$. 
(b) Possible stable location of the 2nd-layer atoms are shown by circles on top of a $a \times a$ parallelogram 
constructed from the 4 neighboring 1st-layer atoms.
(c) Structure of discretized lattice for the 2nd-layer model. Lattice points are shown by circles.
}
\label{fig:lattice}
\end{figure}

In the 4/7 phase, the 3/4 of $^3$He atoms on the 2nd layer occupy points just above mid points of the edges of triangles formed by the 1st-layer atoms whereas the 1/4 occupy points just above  the 1st-layer atoms in a regular fashion as shown in Fig.~\ref{fig:lattice}(a). 
Here, open circles represent the atoms on the 1st layer and 
shaded circles represent actual locations of $^3$He atoms on the 2nd layer when solidified. 
%If $^3$He atoms are used for the 1st-layer, the saturated 1st-layer density is given by 
%$\rho_{1}=0.064$ atom/\AA$^2$ and then the lattice constant 
%for the triangular lattice of the 1st layer is $a=3.1826$ \AA~\cite{Elser}.  
If $^3$He atoms are adsorbed on the 1st layer, it forms a triangular lattice with the lattice constant $a=3.1826$ \AA 
at the saturation density $\rho_{1}=0.114$ atom/\AA$^2$~\cite{Elser}.
%As seen in Fig.~1(a), the 4/7 phase is regarded as the solidification of the $^3$He atoms by the interatomic interaction. 
%This view allows to draw a unit cell of the translational symmetry-broken state 
%as the area indicated by the solid line 
%in Fig.~1(a).

The location of the 2nd-layer atoms is in principle determined as stable points in continuum space. In the present treatment, we simplify the continuum by discretizing it with as much as large number of lattice points kept as candidates of the stable points in the solid.
To illustrate the discretization, we cut out from Fig.~\ref{fig:lattice}(a) a parallelogram whose corners are just above 4 atoms on the 1st layer as in Fig.~\ref{fig:lattice}(b). 
Possible stable locations of $^3$He atoms on the 2nd layer 
are (1) the points just above the mid points of the 1st-layer atoms, (2) 
the centers of the regular triangles and (3) the points just above the 1st-layer atoms. 
Therefore, we employ totally 6 points as the discretized lattice points
in a parallelogram as circles in Fig.~\ref{fig:lattice}(b). 
Since a unit cell in Fig.~\ref{fig:lattice}(a) contains 7 parallelograms, 
it contains 42 lattice points in total as illustrated as circles in Fig.~1(c). 
Now the 4/7 solid phase is regarded as a regular alignment of 4 atoms on 42 available lattice points in the unit cell shown in Fig.~1(c).
%The possible location for $^3$He on the 2nd layer above the 1st layer is 
%illustrated as in Fig.1, where all the equivalent sites to the A site and B site 
%are considered. 
%
%%%%%%%%%%%%%%%%%%%  Fig.1  %%%%%%%%%%%%%%%%%%%%%%%%%%%%%%%%%%%%%%%%%%%%%%%%%%
%%%%%%%%%%%%%%%%%%%%%%%%%%%%%%%%%%%%%%%%%%%%%%%%%%%%%%%%%%%%%%%%%%%%%%%%%%%%%%

\begin{figure}[t]
\begin{center}
\includegraphics[width=7.8cm]{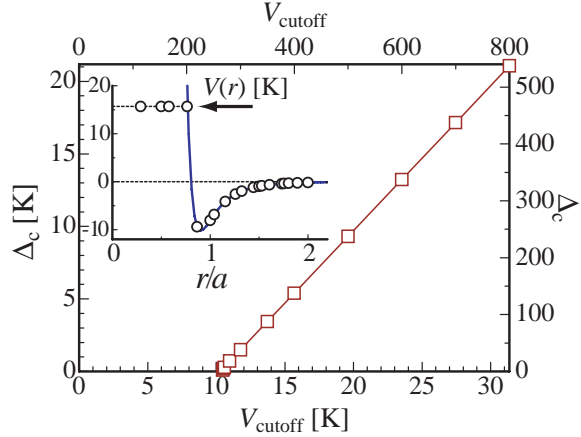}
\end{center}
\caption{
(Color online) 
%The 
$V_{\rm cutoff}$ dependence of the ``charge gap" $\Delta_{\rm c}$. 
The inset shows the He-He interaction $V(r)$ vs. $r$ (see text).
}
\label{fig:VLJ}
\end{figure}

We employ the Lennard-Jones potential 
\begin{eqnarray}
V_{\rm LJ}(r)=4\epsilon \left[
\left(
\sigma/r
\right)^{12}
-
\left(
\sigma/r
\right)^{6}
\right],
\label{eq:LJ}
\end{eqnarray}
for the inter-helium interaction%
where $\epsilon=10.2$ K and $\sigma=2.56$ \AA~\cite{Boer}.
More refined Aziz potential is expected to give similar results under this discretization. 
In the inset of Fig.~\ref{fig:VLJ}, $V_{\rm LJ}(r)$ vs. 
$r$ in the unit of $a$ is shown by the bold solid curve. 
The interaction term of the lattice model is given by $H_{V}=\sum_{ij}V_{ij}n_i n_j$ 
($n_i$ is a number operator of a Fermion on the $i$-th site) 
with $V_{ij}$ taken from the spatial dependence of eq.~(\ref{eq:LJ}) on the lattice points. 
In the actual $^3$He system, the chemical potential of the 3rd layer is estimated to be 16 K higher than 
the 2nd layer~\cite{Whitlock}. 
%参考文献Whitlock et al.PRB58(98) 8704?). 
$^3$He atoms may fluctuate into the 3rd layer over this chemical potential difference and  
it is signaled by an increase of the specific heat for $T>1$ K~\cite{tuji,Sciver,Graywall}
%(ここでvan Sciver and Vilches PRB 18(78) 285とGreywall PRB41(90) 1842も引用すべき), 
as is reflected by the entropy per site larger than $k_{\rm B}\log 2$. 
To take account of this density fluctuation, we here mimic the allowed occupation on the 3rd-layer by introducing a simple finite cutoff $V_{\rm cutoff}$  
for $V_{ij}$ within the same form of Hamiltonian: When $V_{\rm LJ}(r_{ij})$ for $r_{ij}\equiv |{\bf r}_i-{\bf r}_j|$ exceeds $V_{\rm cutoff}$, we take $V_{ij}=V_{\rm cutoff}$ and 
otherwise $V_{ij}=V_{\rm LJ}(r_{ij})$. This allows taking account qualitative but essential part of possible occupation on the 3rd layer by the atoms overcoming $V_{\rm cutoff}$. 
We show the case of $V_{\rm cutoff}=16$ K as indicated by an arrow in the inset of Fig.~\ref{fig:VLJ}. 
Here, the open circles show $V(r_{ij})$ 
on the lattice sites in Fig.~\ref{fig:lattice}(c) for $r/a \le 2$. 
%Since $V_{\rm LJ}(r_{ij})$ within the 4th nearest neighbor pairs is larger than 16 K, we replace it with $V(r_{ij})=V_{\rm cutoff}$. 
%In this way, we take into account the density fluctuation between the 2nd and the 3rd layers. 

%%%%%%%%%%%%%%%%%%%  Fig.2  %%%%%%%%%%%%%%%%%%%%%%%%%%%%%%%%%%%%%%%%%%%%%%%%%%
%%%%%%%%%%%%%%%%%%%%%%%%%%%%%%%%%%%%%%%%%%%%%%%%%%%%%%%%%%%%%%%%%%%%%%%%%%%%%%

Our Hamiltonian for the lattice model 
$H=H_{\rm K}+H_{\rm V}$ consists of the kinetic energy
$H_{\rm K}=-\sum_{\langle ij\rangle}(t_{ij}c_{i}^{\dagger}c_j +{\rm H.C.})$ and $H_V$. 
By using the unit-cell index $s$ and the site index $l$ in the unit cell, 
we have ${\bf r}_i={\bf r}_s+{\bf r}_l$. 

After the Fourier transform, 
$c_{i}=c_{s,l}=\sum_{\bf k}c_{{\bf k},l}e^{{\rm i}{\bf k}\cdot{\bf r}_s}/\sqrt{N_{\rm u}}$, 
the mean-field (MF) approximation with the diagonal order parameter 
$\langle n_{{\bf k},l}\rangle$ leads to
$H_{\rm V}{\sim}H_{\rm V}^{\rm MF}$
\begin{eqnarray}
%H_{\rm V}\sim H_{\rm V}^{\rm MF}
=
\frac{1}{N_{\rm u}}
\sum_{l,m=1}^{42}\sum_{s'}V^{lm}(s')
\sum_{{\bf k}, {\bf p}}
\left[
\langle n_{{\bf k},l} \rangle n_{{\bf p},m}
%\right.
%\nonumber
%\\
%& &
-
%\left.
\frac{1}{2}
\langle n_{{\bf k},l} \rangle \langle n_{{\bf p},m}\rangle
\right], 
\nonumber
\end{eqnarray}
where the inter-atom interaction is expressed as 
$V_{ij}=V_{st}^{lm}=V^{lm}(s')$ 
with ${\bf r}_{s'}={\bf r}_{s}-{\bf r}_{t}$. 
Then, we have the MF Hamiltonian 
$H_{\rm MF}=H_{\rm K}+H_{\rm V}^{\rm MF}$. 
By diagonalizing the $42 \times 42$ Hamiltonian matrix 
for each ${\bf k}$, 
we obtain the energy bands 
$
H_{\rm MF}=\sum_{\bf k}\sum_{l=1}^{42}
E_{l}({\bf k})c_{{\bf k},l}^{\dagger}c_{{\bf k},l}. 
$
%

%As seen in the inset of Fig.~2, $|V(r)|$ decreases as the inter-atom distance increases. 
Here we show the results 
by taking account of the transfers and interactions 
for $|r_{ij}|/a \le 2$ as indicated by the open circles in Fig.~\ref{fig:VLJ}.
%Then, the 19-th neighbor inter-atom interactions $V_{ij}$
%and transfer integrals $t_{ij}$ are retained. 
Then, $V_{ij}$ and $t_{ij}$ for the  $ij$ pairs up to the shortest-19th 
$r_{ij}$ 
are retained. 
For the kinetic energy, several choices of $t_{ij}$ are examined 
and here we show the result for $t_{ij}=t_0/r_{ij}^2$, assuming that 
it is proportional to ${\hbar}/(2m r_{ij}^2)$. 
We note that 
the kinetic energy per atom 
for the 4/7 phase is estimated as 20 K 
by the path-integral Monte Calro (PIMC) simulation~\cite{PIMC}. 
Hence, we evaluate $t_{0}$ 
by imposing the condition, 
$\sum_{\langle ij \rangle}
t_{0}/r_{ij}^2 \langle c_{i}^{\dagger}c_{j}+{\rm H.C.}\rangle/(4N_{\rm u})=20$ K. 
We thus obtain $t_{0}=0.0392$ K, 
which is taken as the energy unit. 
The values of $\epsilon$ and $\sigma$ in eq.~(\ref{eq:LJ}) 
are given by $\epsilon/t_{0}=260.14$ and $\sigma/a=0.8045$, respectively. 
If $t_0$ is determined so as to reproduce the total kinetic energy of 
the PIMC result, the main result measured in the unit of K shown below 
is quite insensitive to the choice of $t_{ij}$~\cite{tij}. 

By solving the MF equations for $H_{\rm MF}$, we have the solution 
of the $\sqrt{7}\times\sqrt{7}$ commensurate structure 
shown in Fig.~\ref{fig:lattice}(a) 
for $V_{\rm cutoff}\ge 267 t_0\equiv V_{\rm cutoff}^{\rm c}$. 
The ``charge gap" opens for $V_{\rm cutoff}\ge V_{\rm cutoff}^{\rm c}$, 
as shown in Fig.~\ref{fig:VLJ}. 
Here, the ``charge gap" is defined by 
$\Delta_{\rm c}=E^{\rm min}_{5}({\bf k})$-$E^{\rm max}_{4}(\bf k)$, 
where $E^{\alpha}_{l}({\bf k})$ denotes the minimum or maximum value of 
the $l$-th band from the lowest. 
The left (right) and bottom (top) axes represent $\Delta_{\rm c}$ and $V_{\rm cutoff}$ 
in the unit of K $(t_0)$, respectively. 
From the specific heat data, $\Delta_{\rm c}$ is estimated 
to be $\sim 1$ K. This leads to $V_{\rm cutoff}/t_0\sim 300$ (namely, 12 K), which is consistent with the chemical potential difference of the 3rd layer $\sim 16$ K~\cite{Whitlock}. 
%再びWhitlock　をrefer.
Since $V_{\rm cutoff}/V_{\rm cutoff}^{\rm c}\sim 1.1$, 
the 4/7 phase is located close to the fluid-solid boundary. 
%We have confirmed that the potential difference for the 2nd-layer atom 
%with the magnitude of 3 K~\cite{Boato} 
The effect of 3 K higher potential on top of the 1st layer $^3$He than other lattice points of the 2nd layer~\cite{Boato} 
merely shifts the $\Delta_{\rm c}$-$V_{\rm cutoff}$ line 
toward larger $V_{\rm cutoff}$: $V_{\rm cutoff}^{\rm c}$ is 
changed from $\sim 11$ K to $\sim 14$ K and hence 
the above conclusion does not change.

\begin{figure}
\begin{center}
\includegraphics[width=7cm]{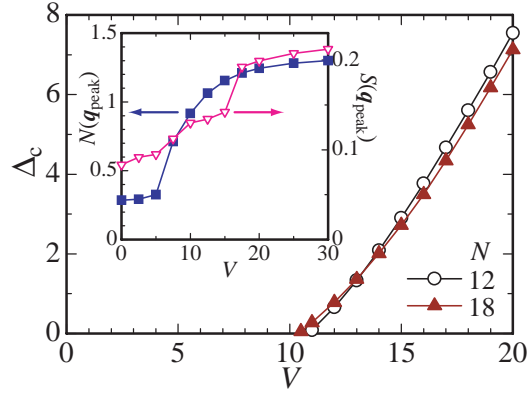}
\end{center}
\caption{
(Color online) 
$V$ dependence of ``charge gap" 
on $N=12$ (open circle) and $N=18$ (filled triangle) triangular lattices 
for $t_1=t_2=t_3=1$ and $V=U$ at $n=1/3$.  
The inset shows $V$ dependence of the peak value of $N({\bf q})$ (filled square) 
and $S({\bf q})$ (open triangle) for $N=12$ under periodic b.c. 
}
\label{fig:cgp}
\end{figure}

\begin{figure}
\begin{center}
\includegraphics[width=5.5cm]{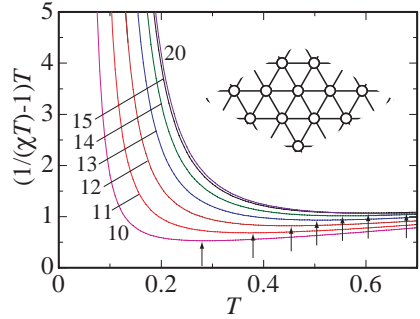}
\end{center}
\caption{
(Color online) 
Temperature and $V$ dependences of susceptibility 
on $N=12$ triangular lattice at $n=1/3$
for $t_1=t_2=t_3=1$ and $U=V$. 
The inset shows a triangular lattice with $N=12$ sites.
}
\label{fig:sus_T}
\end{figure}

To further understand the nature of the solid 
near the fluid-solid boundary, we next consider 
a minimal model 
$\tilde{H}=\tilde{H}_{\rm K}+\tilde{H}_{\rm U}+\tilde{H}_{\rm V}$
with 
$
\tilde{H}_{\rm K}
=
-
\sum_{\sigma}\sum_{\langle ij \rangle}
\left(
t_{ij}
c^{\dagger}_{i\sigma}c_{j\sigma}
+{\rm H.C.}
\right), 
$
$
\tilde{H}_{\rm U}=U\sum_{i}n_{i\uparrow}n_{i\downarrow}
$ 
and 
$
\tilde{H}_{\rm V}
=\sum_{\langle ij \rangle}
V_{ij}
\sum_{\sigma, \sigma'}
n_{i\sigma}n_{j\sigma'}
$, 
%
%\begin{eqnarray}
%H&=&\sum_{\sigma}\sum_{\langle ij \rangle}
%\left(
%t_{ij}
%c^{\dagger}_{i\sigma}c_{j\sigma}
%+{\rm H.C.}
%\right)
%+U\sum_{i}n_{i\uparrow}n_{i\downarrow}
%\nonumber
%\\
%&+&\sum_{\langle ij \rangle}
%V_{ij}
%\sum_{\sigma, \sigma'}
%n_{i\sigma}n_{j\sigma'}, 
%\label{eq:HUV}
%\end{eqnarray}
%%
where $n_{i\sigma}=c^{\dagger}_{i\sigma}c_{i\sigma}$ 
and $\langle ij \rangle$ denotes the pair of the sites. 
To simulate the quantum phase transition between fluid and commensurate solid, 
we consider a triangular lattice with $N=12$ sites 
with $N_{\rm e}=4$ Fermions (see inset of Fig.~\ref{fig:sus_T}). 
When the nearest neighbor repulsion $V \equiv V_{ij}$ is large 
in comparison with the transfer, a commensurate solid 
is expected to be realized.
To make accurate estimates of physical quantities 
we employ the exact diagonalization. 
Here the transfer integrals 
with the $\alpha$th nearest-neighbor $t_\alpha$ for $\alpha \le 3$ 
and the nearest-neighbor repulsion $V$ are retained. 
We take
$t_{1}=t_{2}=t_{3}=1$ and $U=V$ 
to express the large kinetic energy and the effect of $V_{\rm cutoff}$ 
for $^3$He. 
Figure~\ref{fig:cgp} shows the ``charge gap". 
Here, we calculate the ground-state energy 
by introducing the phase factor for the transfer integral: 
$t_{ij}=\tilde{t}_{ij}\exp[i\vec{\phi}\cdot({\bf r}_{i}-{\bf r}_{j})]$, 
where $\vec{\phi}=\phi_1{\bf b}_1+\phi_2{\bf b}_2$ 
with ${\bf b}_i$ being a reciprocal lattice vector which satisfies 
${\bf b}_i\cdot{\bf a}_j=\delta_{ij}$. 
To reduce the finite-size effects, 
the ``charge gap" is defined by 
$\Delta_{\rm c}\equiv{\rm max}\{\mu_{\rm min}^{+}-\mu_{\rm max}^{-}, 0\}$, 
where $\mu_{\rm min}^{+}={\rm min}_{\phi}[E(N_{\rm e}+2)-E(N_{\rm e})]/2$ 
and $\mu_{\rm max}^{-}={\rm max}_{\phi}[E(N_{\rm e})-E(N_{\rm e}-2)]/2$ 
with $E$ being the ground-state energy~\cite{Koretune2007}.
%We take $\phi_{\xi}=m\pi/8$ with $\xi=x, y$ and the integer $m$ running from 0 to 8, i.e. totally 81 mesh points
We take $\phi_{\xi}=\gamma\pi/8$ with $\xi=x, y$ and the integer $\gamma$ running from 0 to 8, i.e., totally 81 mesh points
for $N=12$ and $N=18$ at the filling $n=N_{\rm e}/N=1/3$.
The results show little system-size dependence indicating 
the metal-insulator transition at $V=V_{\rm c}\sim 10$ in the bulk limit. 
The inset in Fig.~\ref{fig:cgp} shows 
the $V$ dependence of the peak value of the equal-time charge and spin 
correlation functions with the periodic boundary condition (b.c.), 
$(\phi_x, \phi_y)=(0,0)$, i.e., 
$N({\bf q})=\sum_{i,j}\exp[{\rm i}{\bf q}\cdot({\bf r}_i-{\bf r}_j)]$
$(\langle n_{i}n_{j}\rangle - \langle n_i \rangle \langle n_j \rangle)/N$ 
and 
$S({\bf q})=\sum_{i,j}\exp[{\rm i}{\bf q}\cdot({\bf r}_i-{\bf r}_j)]
\langle {\bf S}_i\cdot{\bf S}_j
\rangle/(3N)$. 
The peak of $N({\bf q})$ at $(q_x,q_y)=(2\pi/3, 2\pi/\sqrt{3})$ 
increases rapidly around $V=V_{\rm c}$. 
%The peak in $S({\bf q})$ at $(q_x,q_y)=(\pi/3, \pi/\sqrt{3})$ 
%increases at a higher $V_{\rm s}>V_{\rm c}$. 
%These indicate that the charge-ordered insulating state 
The peak in $S({\bf q})$ at $(q_x,q_y)=(\pi/3, \pi/\sqrt{3})$
jumps at a higher $V_{\rm s}>V_{\rm c}$ suggesting that
a commensurate solid for $V>V_{\rm c}$ is stabilized without a spin order for $V<V_{\rm s}$ implying the QSL for $V_{\rm c}<V<V_{\rm s}$.
The realistic choice of $V/V_{\rm c}\sim 1.1$ inferred from the MF study is located in this QSL region.
%We note that for $V/V_{\rm c}\sim 1.1$ 
%$\langle n_{i\uparrow}n_{i\downarrow}\rangle=0.018$ 
%and $\langle n_{i}n_{j}\rangle=0.015$ for the nearest-neighbor bond 
%give a rough estimate for the ratio of the 3rd-layer promotion to the 2nd-layer density.
We note that the ratio of the 3rd-layer promotion to the 2nd-layer density 
is estimated to be about $10$-$20\%$ from the results of double occupancy and 
the nearest-neighbor $\langle n_{i}n_{j}\rangle$ averaged over the 81 phase factors in $N=12$ 
for $V/V_{\rm c}\sim 1.1$.

%%%%%%%%%%%%%%%%%%%  charge gap  %%%%%%%%%%%%%%%%%%%%%%%%%%%%%%%%%%%%%
%%%%%%%%%%%%%%%%%%%%%%%%%%%%%%%%%%%%%%%%%%%%%%%%%%%%%%%%%%%%%%%%%%%%%%%%%%%%%%

\begin{figure} 
\begin{center}
\includegraphics[width=8cm]{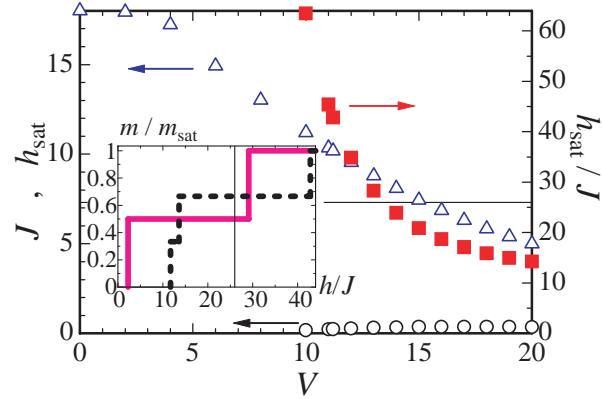}
\end{center}
\caption{
%The $V$ dependence of 
%the exchange interaction $J$ (open circle), 
%the saturation field $h_{\rm sat}$ (open triangle) 
%and its ratio $h_{\rm sat}/J$ (filled square) 
%for $t_1=t_2=t_3=-1$ and $V=U$ at $n=1/3$. 
(Color online) $V$ dependence of
exchange interaction $J$ (open circle),
saturation field $h_{\rm sat}$ (open triangle)
and its ratio $h_{\rm sat}/J$ (filled square)
for $t_1=t_2=t_3=1$ and $V=U$ at $n=1/3$.
The inset shows the magnetization process for $V=11.2$ in $N=12$ (solid bold line)
and $N=18$ (broken line) under periodic b.c. The difference between $N=12$ and 18 
may come from different commensurate structures allowed at $m\sim m_{sat}/2$.
Thin lines (in the inset as well) represent $h/J$ corresponding to 10 Tesla in the experiment when 
we employ $J=0.3$ mK~\cite{Colin}.} 
\label{fig:hsat_V}  
\end{figure}

%To estimate the thermodynamic quantities, we calculate the partition function 
%$ Z={\rm Tr}{\rm e}^{-\beta H}$ in the canonical ensemble by computing all the eigenvalues on the $N=12$ lattice with periodic b.c. 
From the exact diagonalization 
of $N=12$ sites with the periodic b.c., the exchange interaction $J$ is estimated from high-temperature part of $\chi(T)$ 
by the fitting of the high-temperature expansion 
$\chi(T)=(1-3J/T)/T$ 
on the triangular lattice~\cite{chi_HTE}. 
By plotting $(1/(\chi T)-1)T$ vs. $T$ as in Fig.~\ref{fig:sus_T}, 
we estimate $J$ from the flat part indicated by the arrows.
The system-size dependence of $J$ 
is quite small as known in the Hubbard chain~\cite{shiba1}, 
where $\chi(T)$ at high $T$ is determined by the local process. 
Figure \ref{fig:hsat_V} (open circle) shows $J$ for each $V$ obtained in this way.

%%%%%%%%%%%%%%%%%%%  susceptibility  %%%%%%%%%%%%%%%%%%%%%%%%%%%%%%%%%%%%%
%%%%%%%%%%%%%%%%%%%%%%%%%%%%%%%%%%%%%%%%%%%%%%%%%%%%%%%%%%%%%%%%%%%%%%%%%%%%%

The magnetization is calculated 
by adding 
the Zeeman term to $\tilde{H}$: $\tilde{H}-h\sum_{i}S^{z}_{i}$. 
We define the saturation field $h_{\rm sat}$ 
at which the total magnetization $m=\sum_{i}\langle S^{z}_{i}\rangle/N$ reaches 
its saturation value, $m_{\rm sat}=n/2$. 
Figure \ref{fig:hsat_V} shows $h_{\rm sat}$ for the $N=18$ sites (open triangle) 
under the periodic b.c. 
The present saturation field at $V=U=0$ for $N=18$
reproduces the exact
bulk limit $h_{\rm sat}=18.0$,
which is nothing but the width of the $n$-filled band at $h=0$.
%The data reproduces the bulk-limit value at $V=0$. 
%This and the fact that $h_{\rm sat}$ shows the similar $V$ dependence in the $N=12$ sites
%suggest that $h_{\rm sat}$ in Fig.~\ref{fig:hsat_V} is close to the bulk-limit. 
%A remarkable result here is that $h_{\rm sat}$ and hence $h_{\rm sat}/J$ as well increase 
%largely in the insulating phase near the metal-insulator boundary, $V=V_{\rm c}$. 
%This reproduction together with a very similar $V$ dependence of $h_{\rm sat}$ for the $N=12$ sites
%suggest that $h_{\rm sat}$ in Fig.~\ref{fig:hsat_V} is close to the bulk-limit.
%This is one of the central results in this letter: $h_{\rm sat}$ and hence $h_{\rm sat}/J$ as well largely increase
%in the commensurate solid near the solid-fluid boundary, $V=V_{\rm c}$.
%This reproduction together with slightly smaller $h_{\rm sat}$ for all $V$ at $N=12$ sites suggests that $h_{\rm sat}$ in 
%Fig.~\ref{fig:hsat_V} is close to but a lower bound of the bulk-limit.
%This is one of the central results in this letter: $h_{\rm sat}$ and hence $h_{\rm sat}/J$ as well largely increase in the 
%commensurate solid near the solid-fluid boundary, $V=V_{\rm c}$. 
This reproduction together with slightly smaller $h_{\rm sat}$ for $N=12$ (see the inset of 
Fig.~\ref{fig:hsat_V}) suggests that $h_{\rm sat}$ in
Fig.~\ref{fig:hsat_V} is close to the bulk-limit. 
This is one of our central results: $h_{\rm sat}$ and hence $h_{\rm sat}/J$ as well largely increase in the
commensurate solid near the solid-fluid boundary, $V=V_{\rm c}$.
%This enhancement of $h_{\rm sat}/J$ also appears 
%when $t_2=t_3=0$ (not shown). 
%The enhancement becomes even more prominent 
%when $t_2$ is switched on. 
From Fig.~\ref{fig:hsat_V}, we see that 10 Tesla shown as thin lines 
is still below the saturation magnetic field for a realistic choice of $V=11.2$ 
in agreement with the recent experiment~\cite{ishimoto}.

Although the enhancement of $h_{\rm sat}/J$ also appears
at $t_2=t_3=0$ (not shown),
the enhancement is more prominent
when $t_2$ is switched on.
This is understood by 
the perturbation from the large $V(=U)$ limit. 
When $t_2=t_3=0$, $J$ appears first in the 4th order as 
$J_{(4)}=20t_1^4/(3V^3)$, 
whereas
the 2nd-order term $J_{(2)}=4t_2^2/V$ 
and the 3rd-order term 
$J_{(3)}=-10t_1^2t_2/V^2$ 
appear for $t_2\ne 0$. 
%For $t_2<0$, $J_{(3)}$ gives the ferromagnetic (FM) 
%exchange, which gives rise to the cancellation in 
%$J=J_{(2)}+J_{(3)}+J_{(4)}$. 
For $t_2>0$, $J_{(3)}$ becomes ferromagnetic (FM),
which partially cancels $J_{(4)}$ in
$J=J_{(2)}+J_{(3)}+J_{(4)}$.
%This also captures the essence of
%the cancellation among $J_2>0$ antiferromagnetic (AF), $J_3<0$ (FM) and $J_4>0$ (AF) 
%in the MSE with $n$-body exchange interactions $J_n$~\cite{Roger}. 
This is similar to 
the cancellation among antiferromagnetic (AF) $J_2>0$, FM $-J_3<0$ and AF $J_4>0$
in the MSE with $n$-body exchange interactions $(-1)^nJ_n$~\cite{Roger}.
%Our result indeed reproduces the half-magnetization plateau predicted by
%the MSE model~\cite{misguich,momoi} and observed experimentally~\cite{ishimoto}
%(see the inset of Fig.~\ref{fig:hsat_V}).
In short, the enhancement of $h_{\rm sat}/J$ is largely driven by
the density fluctuations near the fluid-solid
boundary, supplemented by the reduction of AF exchange through
partial cancellation by FM MSE.

%%%%%%%%%%%%%%%%%%%  hsat vs V  %%%%%%%%%%%%%%%%%%%%%%%%%%%%%%%%%%%%%
%%%%%%%%%%%%%%%%%%%%%%%%%%%%%%%%%%%%%%%%%%%%%%%%%%%%%%%%%%%%%%%%%%%%%%%%%%%%%%

Let us discuss the significance of the density fluctuations near the fluid-solid boundary in terms of the observed QSL. 
The QSL in $\kappa$-$\rm ET_2Cu_2(CN)_3$~\cite{Shimizu} 
is found
in the region of a tiny charge gap~\cite{Kezmarki}, which consistently reproduces the QSL numerically found near the metal-insulator boundary in the Hubbard model on the triangular lattice~\cite{KI,MWI,Watanabe}. 
The QSL is suppressed when the density fluctuations are suppressed at large $U$~\cite{Mizusaki} consistently with the absence of the QSL phase reported in the spin-1/2 Heisenberg model on the triangular lattice~\cite{Bernu}.  They suggest the importance of density fluctuations for the realization of the QSL in the 4/7 phase of $^3$He.

The density excitations over the energy $\Delta_{\rm c}\sim 1$ K make a peak in the specific heat $C(T)$ at $T\sim 1$ K 
in addition to a low-temperature peak around $T=10^{-1}\sim 1$ mK 
reflecting the spin excitations, 
since the exchange interaction is estimated as 
$J_{(2)}=4t_{ij}^2/V_{\rm cutoff}\sim 5\times 10^{-4}(t_{ij}/t_0)^2$ K. 
The double-peak structure is indeed found in $C(T)$ for $V\ge 10$ in the $N=12$ cluster study 
(not shown) as is observed in $^3$He~\cite{ishida,tuji}.
%The double peak may in fact be observed at smaller $V$ for larger system sizes, which is left for further study.
%The system-size dependence of the low-temperature peak in $C(T)$ and the $T \to 0$ behavior of $C(T)$ is an important future issue to be clarified.

The fluid-solid transition occurs at a very large $V_{\rm cutoff}/t_0\sim 300$
as seen in Fig.~\ref{fig:VLJ}, which reflects the general tendency that the
commensurate solid phase dramatically shrinks when the period of the density order becomes long~\cite{noda}. This explains why the fluid-solid boundary is located near such a large chemical potential difference of the 3rd layer.

%第3層に逃げている粒子の割合についてどこかに書く。

In summary, 
we have shown that the minimal model for $^3$He adsorbed on the graphite should consider the 
density fluctuation to the upper layers. In particular, the properties of the 4/7-solid phase on the 2nd layer are understood only by considering the density fluctuations on the 3rd layer, which makes the real system close to the fluid-solid transition beyond the description by the MSE model. The magnetic field required for the magnetization saturation is largely enhanced in agreement with the experiments. 
The density fluctuations also serve as a key for stabilizing the QSL. Our study predicts that
when the lattice constant of the 1st-layer solid can be changed, the 4/7 solid phase easily changes to fluid.
Experimental tests would be highly desired.

\section*{Acknowledgment}

We thank H. Ishimoto for supplying us with experimental data prior to publication.
This work is supported by Grants-in-Aid for Scientific
Research on Priority Areas under the grant numbers
17071003, 16076212 and 18740191 from MEXT, Japan.
A part of our computation has been done at the Supercomputer Center 
in ISSP, University of Tokyo. 

%\appendix
%\section{Sample}


\begin{thebibliography}{99}
\bibitem{tij} In case of $t_{ij}=-t_0$ 
%for the nearest- to 19th-neighbor transfers, 
for $ij$ pairs up to the shortest-19th $r_{ij}$, 
the result of $\Delta_{\rm c}$ K vs. $V_{\rm cutoff}$ K is nearly the same 
as Fig.~\ref{fig:VLJ}.
\end{thebibliography}
\end{document}